\documentclass{webofc}
\usepackage[varg]{txfonts}   
\newcommand{\krl}{\ensuremath{\kern-0.18em}}
\newcommand{\krr}{\ensuremath{\kern-0.09em}}
\newcommand{\tms}{\ensuremath{\kern-0.1em\times\kern-0.2em}}

\newcommand{\ptt}{\ensuremath{p_{\mathrm{T}}}\xspace}
\newcommand{\pb}{Pb+Pb\xspace}
\newcommand{\ppb}{$p$+Pb\xspace}
\newcommand{\pp}{$p$+$p$\xspace}
\newcommand{\au}{Au+Au\xspace}
\newcommand{\cu}{Cu+Cu\xspace}
\newcommand{\dau}{$d$+Au\xspace}
\newcommand{\ada}{$A$+$A$\xspace}

\newcommand{\rs}[1][7~TeV]{\ensuremath{\sqrt{s}=}~#1\xspace}

\newcommand{\rsnn}[1][2.76~TeV]{\ensuremath{\sqrt{s_{NN}}=}~#1\xspace}
\newcommand{\rsnno}{\ensuremath{\sqrt{s_{NN}}}\xspace}

\newcommand{\gvc}{\ensuremath{\mathrm{GeV}\krl/\krr c}\xspace}

\newcommand{\mvcc}{\ensuremath{\mathrm{MeV}\krl/\krr c^{2}}\xspace}

\newcommand{\pion}{\ensuremath{\pi}\xspace}
\newcommand{\pix}{\ensuremath{\pion^{\pm}}\xspace}

\newcommand{\kx}{\ensuremath{K^{\pm}}\xspace}
\newcommand{\km}{\ensuremath{K^{-}}\xspace}
\newcommand{\kp}{\ensuremath{K^{+}}\xspace}
\newcommand{\kz}{\ensuremath{K^{0}_{\mathrm{S}}}\xspace}

\newcommand{\rh}{\ensuremath{\rho^{0}}\xspace}
\newcommand{\rhm}{\ensuremath{\rho(770)^{0}}\xspace}
\newcommand{\rhpi}{\ensuremath{\rh\krl/\pion}\xspace}

\newcommand{\ks}{\ensuremath{K^{*0}}\xspace}

\newcommand{\ph}{\ensuremath{\phi}\xspace}
\newcommand{\phm}{\ensuremath{\ph(1020)}\xspace}

\newcommand{\ksk}{\ensuremath{\ks\krl/K}\xspace}

\newcommand{\phik}{\ensuremath{\ph/K}\xspace}

\newcommand{\phikm}{\ensuremath{\ph/\km}\xspace}

\newcommand{\pphi}{\ensuremath{p/\ph}\xspace}

\newcommand{\dl}{\ensuremath{\Delta^{++}}\xspace}
\newcommand{\dlm}{\ensuremath{\Delta(1232)^{++}}\xspace}
\newcommand{\dlp}{\ensuremath{\Delta^{++}\kern-0.1em/\krl p}\xspace}

\newcommand{\sigs}{\ensuremath{\Sigma^{*\pm}}\xspace}
\newcommand{\sigsm}{\ensuremath{\Sigma(1385)^{\pm}}\xspace}
\newcommand{\sigsl}{\ensuremath{\Sigma^{*\pm}\kern-0.1em/\krl\Lambda}\xspace}

\newcommand{\ls}{\ensuremath{\Lambda(1520)}\xspace}
\newcommand{\lsl}{\ensuremath{\ls\kern-0.1em/\krl\Lambda}\xspace}

\newcommand{\xs}{\ensuremath{\Xi^{*0}}\xspace}
\newcommand{\xsm}{\ensuremath{\Xi(1530)^{0}}\xspace}
\newcommand{\xsx}{\ensuremath{\Xi^{*0}\kern-0.1em/\Xi^{-}}\xspace}

\newcommand{\mpt}{\ensuremath{\langle\ptt\rangle}\xspace}

\newcommand{\dnc}{\ensuremath{\langle dN_{\mathrm{ch}}\kern-0.06em /\kern-0.13em d\eta\rangle}\xspace}

\newcommand{\raa}{\ensuremath{R_{AA}}\xspace}

\newcommand{\rzz}{\ensuremath{\rho_{00}}\xspace}

\DeclareMathSymbol{\Delta}{\mathalpha}{letters}{"01}
\DeclareMathSymbol{\Lambda}{\mathalpha}{letters}{"03}
\DeclareMathSymbol{\Xi}{\mathalpha}{letters}{"04}
\DeclareMathSymbol{\Sigma}{\mathalpha}{letters}{"06}

\begin{document}
\title{Resonance Production in Heavy-Ion Collisions}

\author{\firstname{Anders G.} \lastname{Knospe}\inst{1}\fnsep\thanks{\email{anders.knospe@cern.ch}}
}

\institute{The University of Houston, Houston, TX, USA
          }

\abstract{%
Hadronic resonances are unique probes that allow the properties of heavy-ion collisions to be studied. Topics that can be studied include modification of spectral shapes, in-medium energy loss of parsons, vector-meson spin alignment, hydrodynamic flow, recombination, strangeness production, and the properties of the hadronic phase. Measurements of resonances in \pp, $p$+$A$, and $d$+$A$ collisions serve as baselines for heavy-ion studies and also permit searches for possible collective effects in these smaller systems. These proceedings present a selection of results related to these topics from experiments at RHIC, LHC, and other facilities, as well as comparisons to theoretical models.
}
\maketitle
\section{Introduction}
\label{sec:intro}

Measurements of hadronic resonances allow the properties of ion-ion collisions to be studied at various stages in their evolution for both small and large collision systems. In these proceedings I will discuss a selection of such measurements. A particular focus will be the ratios of resonance yields to those of stable particles, which can be used to study the properties of the hadronic phase and for which a large set of measurements of many different resonances now exists. I will also discuss the ways in which resonances can contribute to studies of strangeness production and the various mechanisms that determine the shapes of particle \ptt spectra. These proceedings will also touch on the nuclear modification factors, spin alignment, and elliptic flow of resonances and the possibility of modifications to their spectral shapes. In the discussion that follows, masses will usually be omitted from particle symbols: \rhm, $K^{*}(892)$, \phm, \dlm, \sigsm, and \xsm will be denoted as \rh, $K^{*}$, \ph, \dl, \sigs, and \xs, respectively. Except where explicitly stated, the results include both particles and antiparticles. Table~\ref{tab:properties} summarizes the properties~\cite{PDG} of the resonances discussed. Note the variety of masses, lifetimes, and quark contents.

\begin{table}
\centering
\caption{Properties of the resonances discussed~\cite{PDG}. The results discussed in these proceedings were obtained by reconstructing the decays of the resonances in the listed channels. }
\label{tab:properties}
\begin{tabular}{| c | c | r | r | r | l | r |}
\hline
 & Quark & Mass & Width & Lifetime & Decay & Branching \\
Particle & Content &(\mvcc) & (\mvcc) & (fm/$c$) & Mode & Rato (\%) \\\hline
\rh & $(u\bar{u}+d\bar{d})/\krl\sqrt{2}$ & 770 & 150 & 1.3 & $\pi^{-}\pi^{+}$ & 100\\
$K^{*+}$ & $u\bar{s}$ & 892 & 50.3 & 3.92 & $\pi^{+}\kz$ & 33.3\\
\ks & $d\bar{s}$ & 896 & 47.3 & 4.17 & $\pi^{-}K^{+}$ & 66.6\\
\ph & $s\bar{s}$ & 1019 & 4.247 & 46.46 & $K^{-}K^{+}$ & 48.9\\\hline
\dl & $uuu$ & 1232 & 117 & 1.69 & $\pi^{+}p$ & 99.4\\
$\Sigma^{*+}$ & $uus$ & 1383 & 36.0 & 5.48 & $\pi^{+}\Lambda$ & 87\\
$\Sigma^{*-}$ & $dds$ & 1387 & 39.4 & 5.01 & $\pi^{-}\Lambda$ & 87\\
\ls & $uds$ & 1520 & 15.6 & 12.6 & $K^{-}p$ & 22.5\\
\xs & $uss$ & 1532 & 9.1 & 22 & $\pi^{+}\Xi^{-}$ & 66.7\\
\hline
\end{tabular}
\end{table}

\section{Hadronic Phase}
\label{sec:hadronic}

While the yields of long-lived particles are expected to be fixed at chemical freeze out, the yields of resonances may continue to be modified by scattering processes during the hadronic phase until kinetic freeze out~\cite{Bleicher_Stoecker,Markert_thermal,Vogel_Bleicher}. Resonance yields may be regenerated through the pseudo-elastic scattering of hadrons (\textit{e.g.,} $\pi\pi\rightarrow\rho\rightarrow\pi\pi$). When a resonance decays during the hadronic phase, its decay products may under elastic collisions (which can hinder reconstruction of the resonance) or pseudo-elastic scattering through a different resonance state (which removes information about the original resonance); in these cases, the measurable yield of the original resonance is reduced. The final (measurable) resonance yields at kinetic freeze-out will depend on the interplay between regeneration and re-scattering, which is influenced by the temperature and lifetime of the hadronic phase, the resonance lifetime, and the scattering cross-sections of its decay products. Measurements of the ratios of resonance yields to those of long-lived particles can be used, along with models of the hadronic phase, to estimate the temperature and/or lifetime of the hadronic phase~\cite{Markert_thermal,Torrieri_thermal,Torrieri_thermal_2001b}. Experiments at RHIC and the LHC have measured many of the resonances listed in Table~\ref{tab:properties} in various collision systems. The \ks and \ph mesons have been extensively studied, but measurements of the complete set have only become available recently.

\begin{figure}
\centering
\sidecaption
\includegraphics[width=7cm,clip]{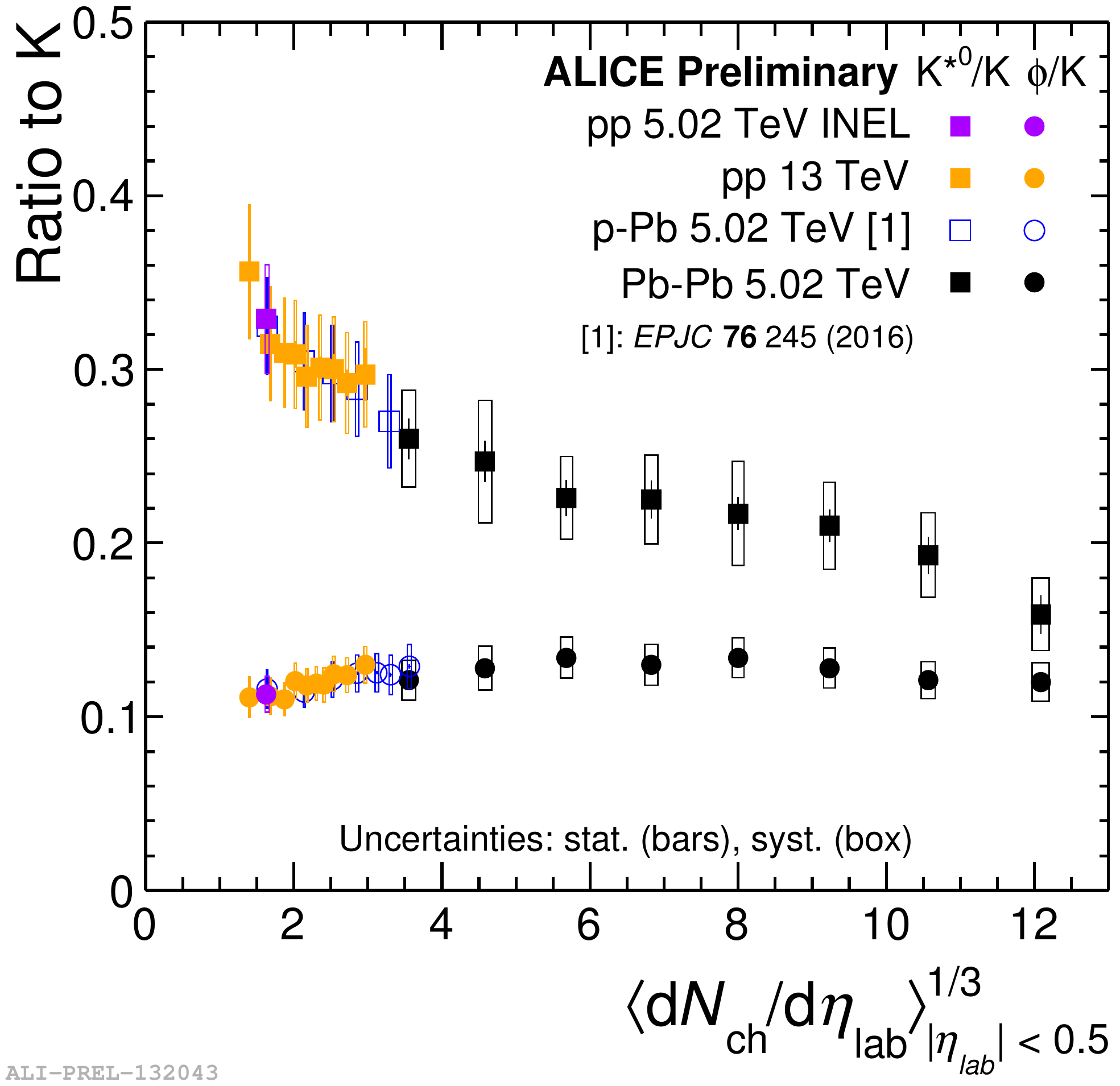}
\caption{Ratios \ksk and \phik as a function of the mean charged particle multiplicity at mid-rapidity in \pp, \ppb, and \pb collisions at \rsnn[5.02~TeV] and in \pp collisions at \rs[13~TeV]~\cite{ALICE_Kstar_phi_pPb5,Neelima_SQM2017_talk,Neelima_SQM2017,Garg_Bergamo_talk}. The yields in the numerator and denominator are the average (not the sum) of particles and antiparticles.}
\label{fig:1}
\end{figure}

Figure~\ref{fig:1} shows the \ksk and \phik ratios measured by the ALICE Collaboration in \pp collisions at \rs[5.02 and 13~TeV] and \ppb and \pb collisions at \rsnn[5.02~TeV] as functions of the mean charged-particle multiplicity at mid-rapidity~\cite{ALICE_Kstar_phi_pPb5,Neelima_SQM2017_talk,Neelima_SQM2017,Garg_Bergamo_talk}. In \pb collisions, the \ksk ratio decreases with increasing collision centrality, while the \phik ratio does not exhibit any strong evolution with centrality. Similar trends are also observed in \pb collisions at \rsnn and in \au and \cu collisions at RHIC energies (plot and references in~\cite{ALICE_Kstar_phi_PbPb}). This behavior may result from the re-scattering of the decay products of the \ks in the hadronic phase, which reduces the measurable yields of that resonance. In contrast, the \ph lives about 10 times longer and should decay primarily after kinetic freeze-out, meaning that its decay daughters should not be affected by re-scattering. This behavior is qualitatively reproduced by the EPOS model~\cite{EPOS_resonances_PbPb}, which incorporates UrQMD to describe the hadronic phase; when EPOS is run with UrQMD turned off, the \ks suppression is not as well described. The PHSD model~\cite{PHSD_2017} can also describe the suppression qualitatively. For \pb collisions at \rsnn, the central \phik ratio is described by thermal-model calculations~\cite{Stachel_SQM2013}, while the \ksk ratio is below the thermal-model values. The measured \ksk ratio can be used along with models to obtain estimates of the lower limit of the lifetime of the hadronic phase~\cite{Markert_thermal,Torrieri_thermal,Torrieri_thermal_2001b}. The same trends are also visible in \pp and \pb collisions and the ratios at a given multiplicity appear to be independent of the collision system. This raises the possibility that re-scattering in the hadronic phase may also be responsible for the suppression of the \ks yield in high-multiplicity \pp and \ppb collisions.

\begin{figure}
\centering
\includegraphics[width=13cm,clip]{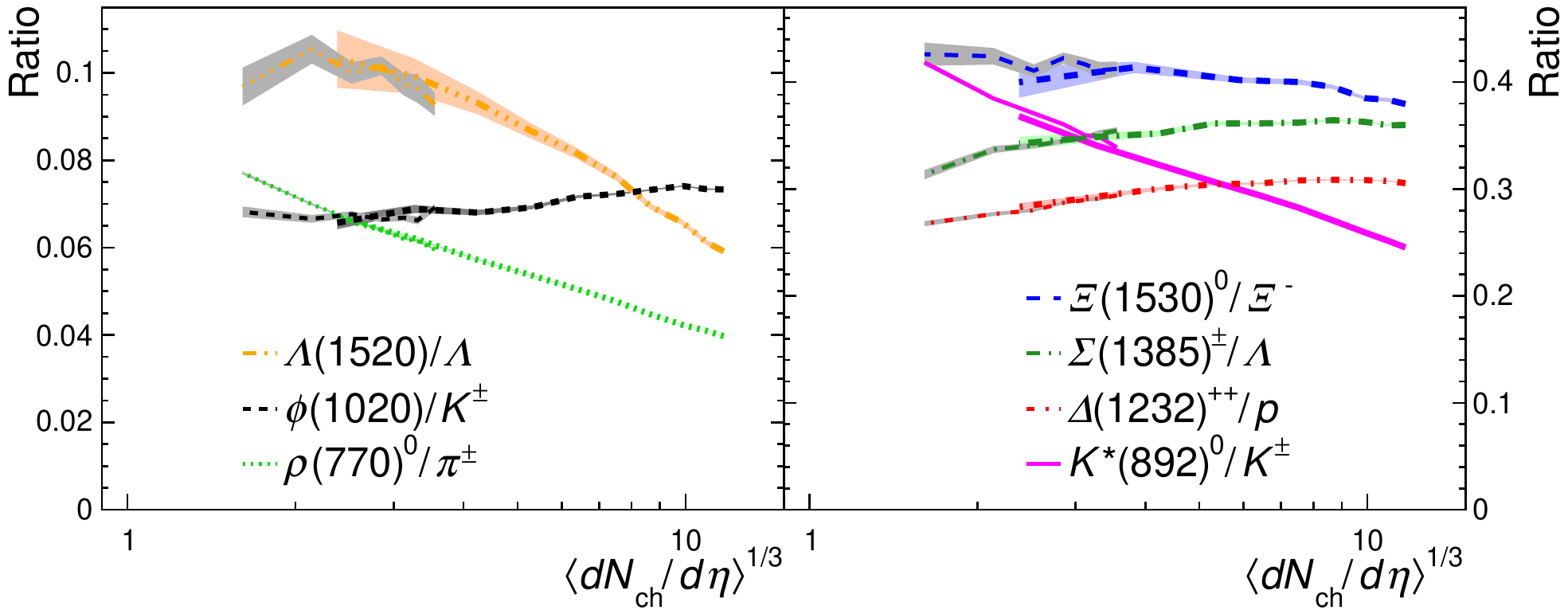}
\caption{Ratios of resonances to long-lived particles given by the EPOS model (version 3.107) for \pb~\cite{EPOS_resonances_PbPb} and \ppb collisions. Shaded bands represent statistical uncertainties. The values shown here qualitatively describe the multiplicity evolution of the ratios measured by the STAR and ALICE collaborations, with the possible exception of the \xsx ratio. The yields in the numerator and denominator are always the sum (not the average) of particles and antiparticles.}
\label{fig:2}
\end{figure}

Similar studies can be performed for other resonances. The \rhpi ratio decreases with centrality in \pb collisions at \rsnn~\cite{Knospe_SQM2016} and is below the thermal model value in central \pb collisions. In contrast, there is no visible suppression of the \rhpi ratio in collisions at \rsnn[200~GeV]~\cite{STAR_resonances_2008}, though at this energy the STAR Collaboration only measured \pp, \dau, and peripheral \au collisions. (The STAR values are also inconsistent with those measured by ALICE at higher energy.) The STAR Collaboration also measured the \dlp ratio in \pp and \dau collisions at \rsnn[200~GeV] and observed no visible suppression trend~\cite{STAR_resonances_2008}. Measurements of the \sigsl ratio in various collision systems at RHIC~\cite{STAR_resonances_2008} and LHC~\cite{ALICE_Sigmastar_Xistar_pPb5} energies do not indicate any significant dependence on energy or system size; the values in central \ada collisions are consistent with thermal-model values. The ALICE Collaboration has measured the \lsl ratio in \pp, \ppb, and \pb collisions~\cite{Song_QM2017,Neelima_SQM2017_talk,Neelima_SQM2017}. The values of this ratio are independent of multiplicity in \pb collisions, while a decreasing trend with centrality is observed in \pb collisions. The measured values in central \pb collisions are below the thermal-model values. The ALICE measurements, including the suppression trend, are consistent with earlier measurements from STAR~\cite{STAR_resonances_2008}. The situation is complicated for the \xsx ratio. No significant centrality dependence in this ratio is observed in \pb collisions at \rsnn~\cite{Song_QM2017,Neelima_SQM2017_talk,Neelima_SQM2017}, but the mid-central and central \pb measurements are lower than the thermal-model values and the values measured in \pp and \ppb collisions~\cite{ALICE_Sigmastar_Xistar_pPb5}. No significant multiplicity dependence is observed for \pp and \ppb collisions. Taken together, these results suggest a possible weak suppression of the \xsx ratio in (mid-)central \pb collisions.  When the hadronic phase is described using UrQMD, EPOS~\cite{EPOS_resonances_PbPb} is able to qualitatively describe the centrality-dependent trends observed for by ALICE for \pb collisions at \rsnn and \ppb collisions at \rsnn[5.02~TeV] (see Figure~\ref{fig:2} for the EPOS values).

In summary, the \rhpi, \ksk, and \lsl ratios are suppressed in central \ada collisions, while the \phik and \sigsl ratios are not. The \xsx ratio may exhibit a weak suppression in \ada collisions, while the \dlp ratio in not suppressed in \dau collisions. It should be noted that the suppression of these resonances is not just dependent on their lifetimes. The \sigs is not suppressed even though its lifetime is similar to that of the \ks; EPOS predicts no suppression for the the \dl in \pb collisions, even though its lifetime is similar to that of the \rh. It is therefore important to remember that the scattering cross-sections of resonance decay products and competition between regeneration and re-scattering can also be important in determining the final resonance yields.

\section{Strangeness Production}
\label{sec:strangeness}

The production of strangeness, including its possible enhancement in central \ada collisions and/or canonical suppression in (low-multiplicity) \pp collisions, is usually studied using common long-lived hadrons; see~\cite{ALICE_multistrange_pp7mult} for a discussion. Resonances may also contribute to such studies due to the fact that have different masses than their long-lived counterparts and that some, like the \ph, have unique quark content.

The $\Lambda/\pi$, $\Xi/\pi$, and $\Omega/\pi$ ratios have been observed to increase as functions of multiplicity in \pp, \ppb, and \ada collisions~\cite{ALICE_multistrange_pPb5,ALICE_multistrange_PbPb}. The magnitude of the enhancement of these three ratios increases with the strangeness content of the numerator particle, while the $p/\pi$ ratio is essentially constant in \ppb collisions. The fact that the \sigsl and \xsx ratios are constant with multiplicity in \ppb collisions~\cite{ALICE_Sigmastar_Xistar_pPb5} indicates that the \sigs and \xs resonances (which are more massive than the $\Xi$) are enhanced by the same amount as their ground-state counterparts, suggesting that it is indeed the strangeness, and not the mass, that controls the enhancement.

The \ph meson has a unique quark content: it is composed only of strange valence quarks, but has no net strangeness. It is therefore expected that while particles with open strangeness should be affected by the canonical suppression of strangeness production in low-multiplicity collisions, the \ph should be unaffected~\cite{Kalweit_strangeness_2016}. In \pp and \ppb collisions, the \phik ratio has been observed to be fairly constant, with a possible weak increase with multiplicity~\cite{Neelima_SQM2017_talk,Neelima_SQM2017}. Similarly, the $\Xi/\ph$ ratio exhibits a possible weak increase with multiplicity. In the context of these measurements, the \ph therefore seems to behave like a particle with 1 to 2 units of strangeness. Why the evolution of \ph yields seems to follow the evolution of canonically suppressed particles is an open question which requires more input from the theoretical community.

The \phik ratio has been measured in \ada collisions by multiple experiments from $\sim2$~GeV to 13~Tev~\cite{HADES_phi_2017,ALICE_Kstar_phi_PbPb}. The ratio is essentially constant over most of that range, but an increase is seen for $\rsnno<4$~GeV. This increase can be explained by statistical models with a strangeness correlation radius of $R_{\mathrm{C}}\approx2.2$~fm: strangeness in required to be conserved within this radius, which would suppress kaon yields but not affect the \ph yields. It can also be noted that the \phikm ratio approaches 0.5 at these low energies; this indicates a sizable feed-down contribution to the \km yield and could explain the different slopes observed in the \km and \kp \ptt spectra.

\section{Shapes of $\boldsymbol{\ptt}$ Spectra}

Resonances, the \ph in particular, can also be useful for studies of the mechanisms that determine the shapes of hadron \ptt spectra. When compared to measurements in \pp collisoions, baryon-to-meson ratios, such as $p/\pi$ and $\Lambda/\kz$, exhibit a centrality-depended enhancement for $1.5\lesssim\ptt\lesssim 5$~\gvc in \pb collisions~\cite{ALICE_piKp_PbPb,ALICE_k0s_Lambda_PbPb}. Explanations for this enhancement include hydrodynamic behavior and hadronization through recombination. The hydrodynamic interpretation~\cite{VISH2p1_MCGlb,VISH2p1_MCKLN,KRAKOW} would attribute the enhancement to the differing masses of the numerator and denominator particles, while the recombination interpretation~\cite{Fries_Muller_2003,Coalescence_Review_2008} would attribute it to the differing number of valence quarks (which would lead to differently shaped \ptt spectra for baryons and mesons). In the \pphi ratio, the two particles have very similar masses, permitting the mass and baryon-number effects to be separated. The \pphi ratio is observed to be constant in central \pb collisions for $\ptt<4$~\gvc~\cite{ALICE_Kstar_phi_PbPb}, which is consistent with the simple hydrodynamic picture (although some recombination models can also describe a near-constant behavior~\cite{Minissale_2015}).

\begin{figure}
\centering
\includegraphics[width=7cm,clip]{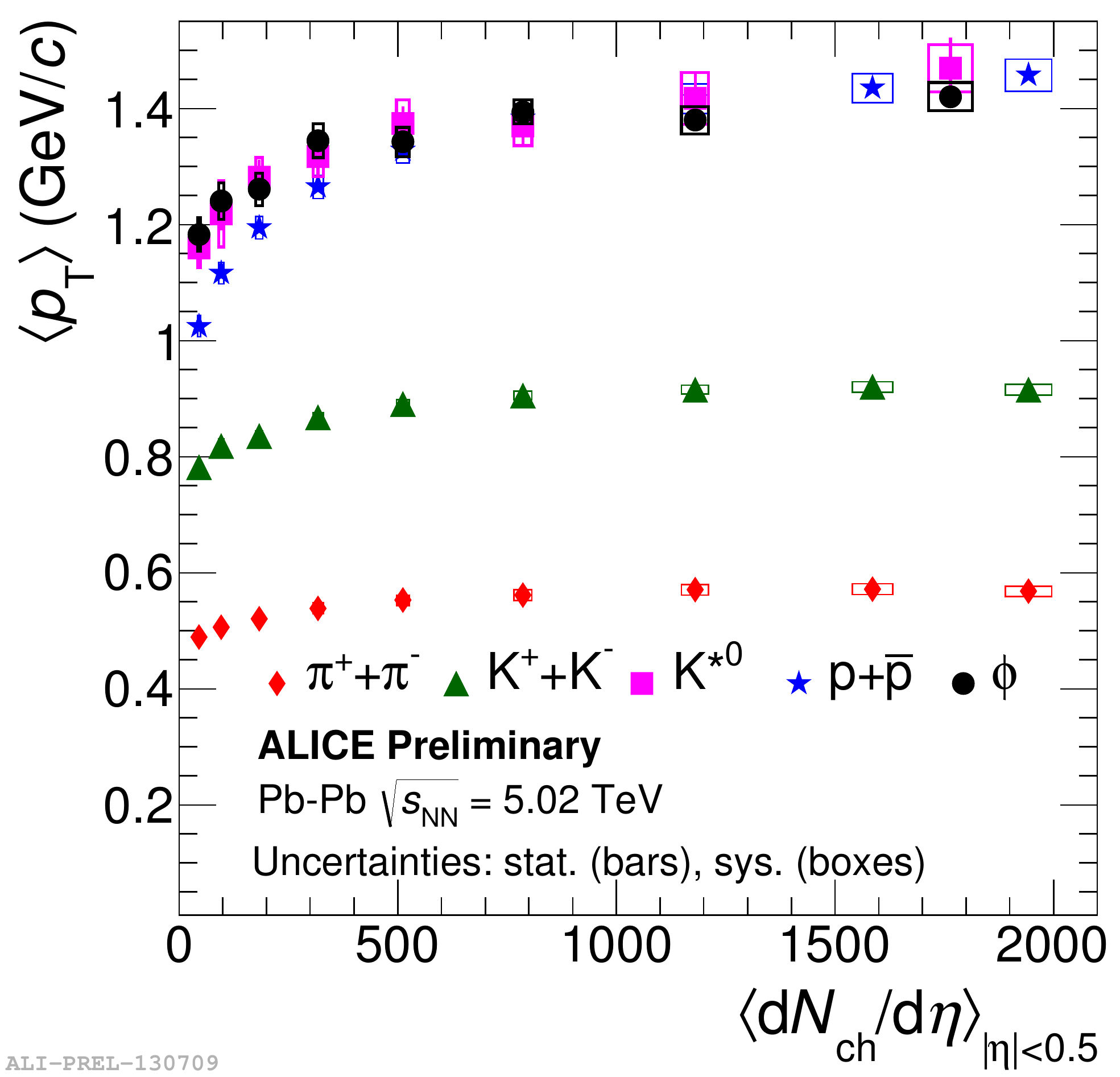}
\includegraphics[width=7cm,clip]{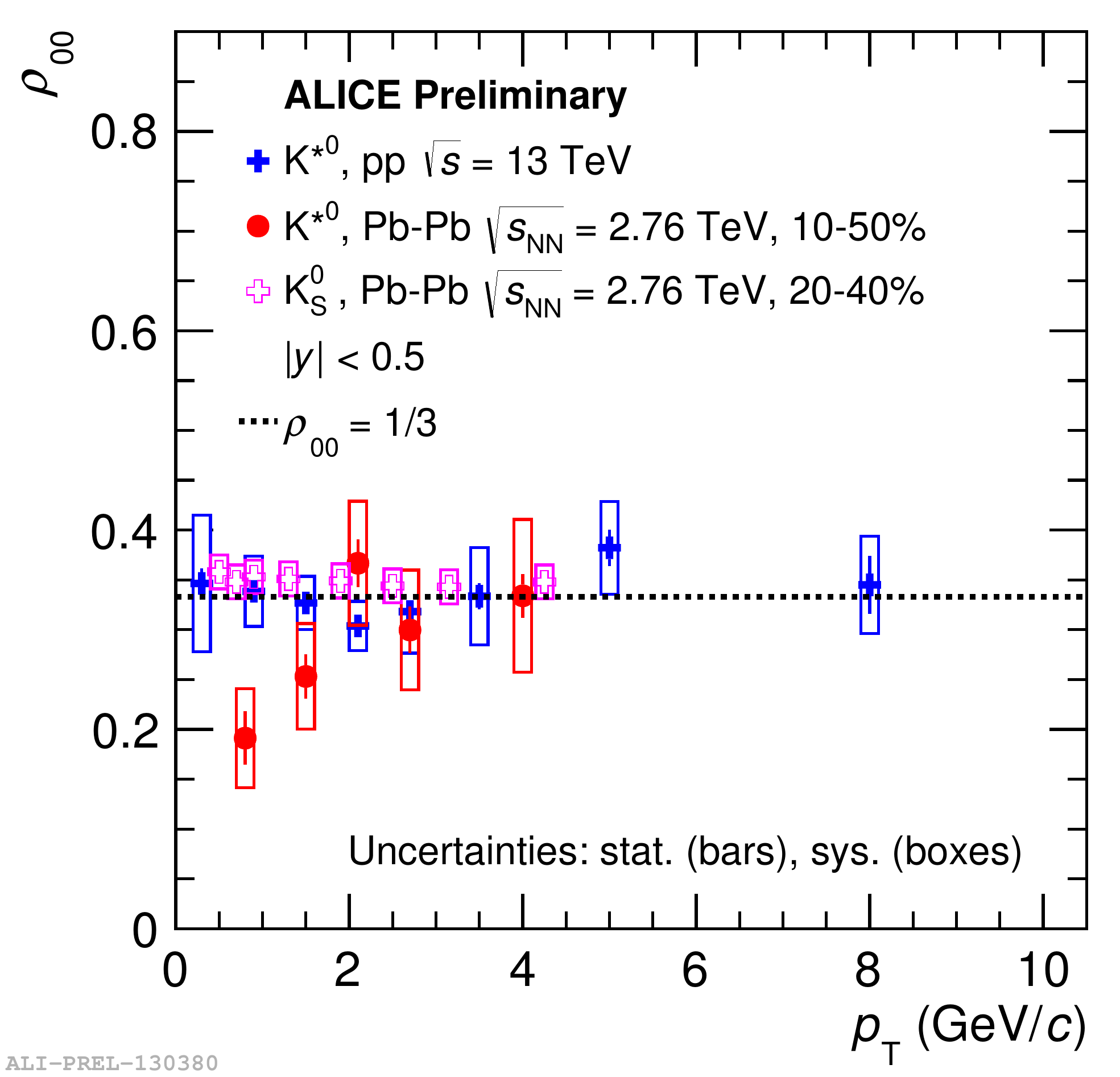}
\caption{\textbf{Left:} \mpt values for \pix, \kx, (anti)protons, \ks, and \ph in \pb collisions at \rsnn[5.02~TeV] for different centrality classes~\cite{Neelima_SQM2017_talk,Neelima_SQM2017}. \textbf{Right:} Spin alignment of the \ks and \kz measured by ALICE.~\cite{Mohanty_SQM2017}}
\label{fig:3}
\end{figure}

The mean transverse momentum \mpt can also be used to study how that shapes of hadron \ptt spectra depend on particle properties, including quark content and mass.
The similarly shaped proton and \ph \ptt spectra in central \ada collisions result in similar \mpt values for these particles~\cite{Neelima_SQM2017_talk,Neelima_SQM2017,ALICE_Kstar_phi_PbPb} (see Figure~\ref{fig:3}a), consistent with a hydrodynamic interpretation.  In central \ada collisions, hadron \mpt values are ordered with particle mass. However, this is not the case for smaller collision systems, in which the \ks and \ph take on larger \mpt values than the long-lived baryons with similar masses~\cite{ALICE_Kstar_phi_pPb5}. The \ph-meson \mpt values even approach those for the $\Xi$ (which has a 30\% greater mass) in \pp and low-multiplicity \ppb collisions. It is not entirely clear whether this violation of mass ordering is due to differences between baryons and mesons or due to the fact that the \ph and \ks are resonances. For all light-flavor hadron species (including resonances), the \mpt values in \ppb collisions follow different trends with multiplicity than the \mpt values observed in \pb collisions. The \ppb values increase faster and in the highest multiplicity collisions can reach (or exceed) the \mpt values seen in central \pb collisions.

\section{Other Topics}

The nuclear modifications factor of the \rh, \ks, and \ph mesons have been measured in \pb collisions at \rsnn~\cite{Knospe_SQM2016,ALICE_Kstar_phi_PbPb_highpT}. For $\ptt\gtrsim 8$~\gvc, all light-flavor hadron species, including these three resonances, are suppressed by a factor of 4 to 5. Indeed, $D$ mesons are also suppressed by the same factor, indicating that the suppression is independent of hadron properties, including mass, baryon number, and quark content (at least for hadrons containing only $u$, $d$, $s$, and $c$ quarks). At intermediate \ptt ($2\lesssim\ptt\lesssim 8$~\gvc), there appears to be a separation between the \raa values of baryons and mesons, with the mesons (even heavy ones like the $D$) suppressed more than baryons. There also appears to be a mass ordering among the mesons, with the amount of suppression decreasing with increasing mass; the three resonances mentioned are consistent with this trend.

Non-central \ada collisions will have a large angular momentum and a large magnetic field. One consequence of this could be an observable spin alignment for vector mesons in heavy-ion collisions. The degree of spin alignment is quantified by the matrix element \rzz, which has an expected value of $\frac{1}{3}$ in the absence of any net spin alignment. The STAR Collaboration measured~\cite{STAR_spin_alignment_2008} \rzz for \ph mesons in \pp and \au collisions at \rsnn[200~GeV] and observed no significant differences between the values for the two collision systems. STAR also observed no significant deviation from $\frac{1}{3}$ for the \rzz values for \ks in \au collisions. The ALICE Collaboration has recently measured~\cite{Mohanty_SQM2017} \rzz for \ks in \pp collisions at \rs[13~TeV] and \pb collisions at \rsnn (see Figure~\ref{fig:3}b). While ALICE observes no significant deviation from $\frac{1}{3}$ for \pp collisions, there are hints of a deviation below $\frac{1}{3}$ (at the 2.5-$\sigma$ level) at low \ptt in non-central (10-50\%) \pb collisions. For comparison, ALICE also measured \rzz for \kz (a pseudoscalar meson) in non-central \pb collisions at the same energy and observed no deviation from $\frac{1}{3}$.

Experiments have also searched for evidence of modifications to the spectral shapes of resonances. Measured excesses of dileptons at low invariant mass in \ada collisions may be due to melting of the \rh meson. The STAR Collaboration has measured~\cite{STAR_resonances_2008} a \ptt-dependent negative mass shift for \rh, $K^{*}$, \dl, and \sigs in \pp and \dau collisions (with a hint of dependence on the collision multiplicity), which may be due to interference, re-scattering, and/or Bose-Einstein correlations involving resonance decay products and other hadrons in the system. The ALICE Collaboration may also observe a hint of a negative mass shift for the \ks meson in \pb collisions~\cite{ALICE_Kstar_phi_PbPb} (although the size of the uncertainties on the measurement prevent strong statements from being made). A recent paper on the PHSD model~\cite{PHSD_2017} explains this shift in terms of medium modification of the \ks spectral function and re-scattering and absorption in the hadronic phase.

Resonances can also be used in studies of elliptic flow. STAR has observed a violation of $v_{2}$ mass ordering for protons and \ph mesons, with $v_{2}(\ph)$ much greater than $v_{2}(p)$ at low \ptt ($\ptt\lesssim 0.7$~\gvc) in \au collisions at \rsnn[200~GeV]~\cite{STAR_v2_2016}. This may be due to hadronic re-scattering affecting the proton for low \ptt. PHENIX~\cite{PHENIX_phi_d_v2_2007} and ALICE~\cite{Bertens_QM2017} have measured $v_{2}$ of the \ph meson as functions of $KE_{\mathrm{T}}$ and \ptt, respectively. ALICE observes that for $\ptt<2$~\gvc, the $v_{2}$ values of the \ph are the same as those of the proton, consistent with the mass ordering expected from hydrodynamic behavior. It is observed that for $KE_{\mathrm{T}}\gtrsim 0.8$~\gvc and $\ptt\gtrsim 2$~\gvc, the $v_{2}$ values of the \ph are consistent with those of other mesons, rather than with the proton. Therefore, there seems to be a transition in the behavior of $v_{2}$ values: with mass ordering observed at low \ptt and baryon-meson splitting seen at intermediate \ptt.

\section{Conclusions}

Resonances are useful in a wide variety of studies in heavy-ion physics. As a function of collision centrality or multiplicity, the yields of \rh, \ks, and \ls are observed to be suppressed; the yields of \ph, \dl, and \sigs are not suppressed; and the \xs may be weakly suppressed. These trends may be explained as consequences of re-scattering of resonance decay products in the hadronic phase and may be used in future studies to learn more about the properties of the hadronic phase. The observed suppression trends (except possibly that of the \xs) are consistent with values obtained from the EPOS model with UrQMD. The \sigsl and \xsx ratios are independent of multiplicity in \ppb collisions, suggesting that the observed increases in strange baryon yields is due to the baryons' strangeness content and not their masses. In small systems, the yields of the \ph meson appear to increase with multiplicity similar to particles with open strangeness, even though the \ph should not be subject to canonical suppression. The proton and the \ph, which have similar masses, have similarly shaped \ptt spectra in central \ada collisions (possibly due to hydrodynamic behavior), but mass ordering among hadrons is violated for smaller systems. Measurements of the nuclear modification factors, $v_{2}$ parameters, spin alignment, and modified spectral functions of resonances may help us further understand the physics of heavy-ion collisions.

\section{Acknowledgements}

I thank the ALICE Collaboration for the use of some of its figures, C. Markert and K. Werner for their help with the EPOS model, and R. Bellwied for making this presentation possible.

\bibliography{refs}

\end{document}